\documentclass[fleqn,usenatbib]{mnras}
\usepackage{newtxtext,newtxmath}

\usepackage[T1]{fontenc}

\DeclareRobustCommand{\VAN}[3]{#2}
\let\VANthebibliography\thebibliography
\def\thebibliography{\DeclareRobustCommand{\VAN}[3]{##3}\VANthebibliography}

\usepackage{color}
\usepackage{booktabs}
\usepackage{enumitem}
\usepackage{hyperref}
\usepackage{verbatim}
\usepackage{amsmath,amstext,mathrsfs}
\usepackage[all]{hypcap} 
\usepackage{afterpage}    
\usepackage{marginnote}
\usepackage{makecell}
\usepackage{subfigure}
\usepackage{xfrac}

\usepackage{ifxetex,ifluatex}
\usepackage{fixltx2e} 
\usepackage{multirow}

\DeclareFontFamily{OMS}{oasy}{\skewchar\font48 }
\DeclareFontShape{OMS}{oasy}{m}{n}{%
         <-5.5> oasy5     <5.5-6.5> oasy6
      <6.5-7.5> oasy7     <7.5-8.5> oasy8
      <8.5-9.5> oasy9     <9.5->  oasy10
      }{}
\DeclareFontShape{OMS}{oasy}{b}{n}{%
       <-6> oabsy5
      <6-8> oabsy7
      <8->  oabsy10
      }{}
\DeclareSymbolFont{oasy}{OMS}{oasy}{m}{n}
\SetSymbolFont{oasy}{bold}{OMS}{oasy}{b}{n}

\DeclareMathSymbol{\smallleftarrow}     {\mathrel}{oasy}{"20}
\DeclareMathSymbol{\smallrightarrow}    {\mathrel}{oasy}{"21}
\DeclareMathSymbol{\smallleftrightarrow}{\mathrel}{oasy}{"24}

\defcitealias{LP00}{LP00}

\graphicspath{{./}{Fig/}}

\title[Is Cattail CNM?]{Is the recently discovered large scale filamentary feature {\it Cattail} in the cold or unstable phase?}

\author[Yuen et.al]{
Ka Ho Yuen,$^{1}$\thanks{kyuen@astro.wisc.edu}
Avi Chen,$^{2}$
Ka Wai Ho,$^{1}$
Alex Lazarian$^{1,3}$
\\
$^{1}$Department of Astronomy, University of Wisconsin-Madison, USA\\
$^{2}$Department of Physics, City College, CUNY, USA\\
$^{3}$Centro de Investigación en Astronomía, Universidad Bernardo O’Higgins, Santiago, General Gana 1760, 8370993, Chile
}

\date{Accepted XXX. Received YYY; in original form ZZZ}

\begin{document}
\label{firstpage}
\pagerange{\pageref{firstpage}--\pageref{lastpage}}
\maketitle
\begin{abstract}
A recent publication (\citealt{Li21}) discovered one of the largest filamentary neutral hydrogen features dubbed {\it Cattail} from high resolution FAST observations that might be a new galactic arm of our own Milky Way. However in the analysis, it was suggested that this neutral hydrogen feature is cold despite having $12km/s$ total linewidth. We evaluate the probability whether the {\it Cattail} is actually cold neutral media via the newly developed Velocity Decomposition Algorithm (\citealt{VDA}) and Force Balancing Model (\citealt{HO21}). We discovered that even with the inclusion of the galactic shear term, the feature is still at the unstable neutral media regime. Moreover, we also discover that the {\it Cattail} is two disjoint features in caustics space, suggesting that the {\it Cattail} might have two different turbulent systems. We check the spectra of the individual system separated via VDA to confirm this argument. We do not exclude the existence of smaller scale cold media being embedded within this structure.
\end{abstract}
\section{Introduction} \label{sec:intro}

The atomic hydrogen cold neutral media (HI-CNM) is one of the most important astrophysical observables. Observations like HI4PI \cite{kh15}, GALFA (\citealt{P18}), THOR-HI (\citealt{B16,W20}), FAST (\citealt{Li21}) have improved our understanding of CNM, including its relation to the underlying molecular phase \cite{KH17}, correlation to magnetic fields (\citealt{H01,Clark15}), and its spatial distribution (\citealt{kk16,kh18}. CNM is ubiquitous in the interstellar media (ISM) for both high latitude (\citealt{Clark14, Clark15})) and galactic plane (\citealt{Soler20}). The linear features of CNM makes it one of the most important B-field probes in modern astronomy, with vast applications from cosmology (\cite{CnH19}) to diffuse molecular cloud studies (e.g., \citealt{Clark14,LYH18}).


Recently a large, {\bf cold} HI filamentary feature was discovered by \cite{Li21} and was dubbed "Cattail". The feature is thought to be either a new galactic arm, or the furthest and largest cold gas filament in our galaxy. It is located near the galactic plane. Its length is either 1.1 kpc or as large as 5 kpc \footnote{\cite{Li21} is not clear on the definition of {\it Cattail}. In the introduction it is stated that {\it Cattail} is 1.1 kpc or as large as 5 kpc. Their table 1, which lists the properties of {\it Cattail} including its 1.1 kpc length refers only to the FAST observation at the tip of {\it Cattail}, i.e. a "small" clump with length 1.1 kpc and width 200 pc. However, starting from page 5, {\it Cattail} is  defined as the  giant filament with length 5 kpc, and the paper concludes that "we suggest two possible explanations for {\it Cattail}: it is a giant filament with a length of ~5 kpc, or part of a new arm in the EOG [Extreme Outer Galaxy].". In this paper, due to the limitation of the resolution from HI4PI, we focus on the entire filament. Notice that the tip of {\it Cattail} has higher density, so we expect the entire filament to have lower density on average.} and its width is 200 pc. It is located 22 kpc away from the galactic center, in the Extreme Outer Galaxy (EOG) with a 5:1 aspect ratio. It was observed using Five-hundred-meter Aperture Spherical radio Telescope (FAST) on August 24 2019. 

{\it Cattail} is believed to be cold.\footnote{Li et.al "H1 filaments are normally cold with typical excitation temperature (50K), corresponding to an isothermal sound speed of approximately 0.65 km/s. Therefore, the turbulence may be supersonic in {\it Cattail}."} It is, however, relatively diffuse ($0.03 cm^{-3}$) and possesses a relatively large linewidth (12 km/s). These characteristics are uncommon for CNM. A typical cold filament at $M_s=3$ and $T=200K$ has a thermal linewidth of $v=4-6 km/s$ along the line of sight (\citealt{kh18}). \cite{Li21} associates the large linewidth with the galactic rotation. \footnote{"... there is no evidence for {\it Cattail}'s velocity coming from the influences of stellar feedback or a tidal stream, the most likely explanation is the Galactic Rotation."} However, even taking the upper estimate of the galactic rotation curve (see \citealt{MR19}), at 22 kpc, its contribution is roughly $300km/s \times (200pc)/22kpc \sim 2.7 km/s$. Combining the turbulent and thermal linewidth there is roughly 4-6 km/s of linewidth discrepancy, if we simply assume the {\it Cattail}'s linewidth is from turbulence, CNM thermal broadening and galactic rotation alone. The 12 km/s linewidth that is seen in {\it Cattail} is, therefore, not consistent with the suggestion that it is CNM. 

In this paper we will take a step back and reanalyse the {\it Cattail} region using the statistics of turbulence via the Velocity Decomposition Algorithm (VDA) and a recently theorized Force Balancing Model \cite{HO21}. We shall show that the {\it Cattail} is not a cold but a transient unstable neutral media (UNM) with significant contributions of turbulence. We will also show that it is likely not a statistically coherent 5 kpc long region. In \S \ref{sec:method} we briefly discuss why VDA is important in classifying CNM and UNM, and the major findings of VDA and Force Balancing Model. In \S \ref{sec:results} we discuss our results. In \S \ref{sec:qual} we argue quantitatively why {\it Cattail} is likely not cold. In \S \ref{sec:discussions} we discuss the importance of {\it Cattail} being UNM, and in \S \ref{sec:conclusion} we conclude our paper. 

\section{Applying the findings of VDA  to {\it Cattail}}
\label{sec:method}

In \cite{VDA} a novel technique was presented to reconstruct the velocity caustics $p_v$\footnote{Velocity caustics represent the velocity contributions to channel maps. Caustics are generally caused by critical points of the phase in differential maps thereby creating a hyper-surface in the space of parameters (See \cite{AR85}. They cause a diverging intensity in the PPV cube which show up as elongated regions in channel maps. These features do not correspond to any physical density.} from each channel based on a linear algebra formalism. The technique, among other things, applied the theoretical results established by \cite{LP00,LP04}. The formalism is known to work in multiphase HI media, and tested both in numerics and observations in \cite{VDA}. To summarize, the density fluctuation $p_d$ and velocity caustics $p_v$ (which together make up velocity channels in PPV cubes) can be obtained from the following formulae with the full PPV cube $p({\bf R},v)$ and the column intensity map $I({\bf R})$:
\begin{equation}
\begin{aligned}
    p_v &= p - \left( \langle pI\rangle-\langle p\rangle\langle I \rangle\right)\frac{I-\langle I\rangle}{\sigma_I^2}\\
    p_d &= p-p_v\\
    p_d &=\left( \langle pI\rangle-\langle p\rangle\langle I \rangle\right)\frac{I-\langle I\rangle}{\sigma_I^2}\\
\end{aligned}
\label{eq:VDA_ld2}
\end{equation}\linebreak
The VDA can determine the relative density-to-velocity fluctuations in HI velocity channel maps. This is illustrated with the $\sigma-v$ and $P_d/P_v$ diagrams. The $\sigma-v$ diagram measures the dispersions of $p_d$ and $p_v$ as a function of line-of-sight velocity and tells us whether a particular channel is velocity or density dominant. The $P_d/P_v$ diagram measures the relative importance of density and velocity fluctuations as a function of spatial scales on the plane of sky and also tells us whether a particular channel is velocity or density dominant in different length scales.

The VDA uses the velocity and density contributions in channel maps to determine whether turbulence is dominant over the thermal contributions. If channel maps are velocity dominant (i.e. $P_d/P_v<1$) then turbulence is dominant and it will have larger contribution to the linewidth. VDA also allows us to evaluate the importance of turbulence via the visual feature in the caustics map: The more turbulent the system, the more fractured the caustic map. 

VDA can distinguish whether the feature is subjected to strong thermal broadening. \cite{VDA} show that in the presence of strong thermal broadening, the channel maps across the linewidth would visually look the same. However, CNM is supersonic in nature and has weak thermal broadening. Thus, when we are examining velocity channels whose width is thinner than the turbulent linewidth, the CNM features will look different in different channels, while the strongly thermal broadened features (UNM,WNM) will look the same across the channels. The VDA, therefore, allows us to distinguish whether the visual features in the channel maps are cold or not.

Recently \cite{HO21} proposed a Force Balancing Model that predicts the aspect ratio of the CNM by considering the force balances between the UNM and CNM. Their result showed that the aspect ratio of {\it dynamically stable} CNM in typical turbulence ($M_s\approx 3$, $M_A \approx 0.5$) and density conditions for far-away arms ($n_{CNM,18kpc}\approx 1 cm^{-3}$, $n_{UNM,18kpc} \approx 0.1cm^{-3}$, \citealt{Wolfire03} ) should be at the order of unity, which is significantly smaller than the currently circulating estimate using the \citeauthor{GS95} scaling, (\citeyear{GS95}, see, e.g. \cite{2019ApJ...878..157X}). Notice that if the observed CNM is longer than the theoretical estimate from \cite{HO21}, it is dynamcially unstable and will soon fragment into shorter clumps. \cite{HO21} suggests a new type of instability generated from an attractive pressure force from the UNM. WNM are subjected to \citeauthor{GS95} turbulence since critical balance holds in WNM. However, during the formation of CNM, the pressure balance between the UNM attractive pressure force and the CNM repulsive pressure force along the perpendicular direction of the magnetic field will restrict the length and width of the newly formed CNM. From \cite{HO21}:
\begin{equation}
\text{Aspect ratio} \approx \frac{M_s^2}{M_A} [25.6  \frac{l_\bot}{l_{U,\bot}} - 2(\frac{n_{peak}}{n_C}-1) ]^{-1}
\label{eq:aspect_UNM_compression}
\end{equation}
where $M_{s,A}$ are the sonic and Alfvenic Mach numbers, $l_\bot$ is the perpendicular CNM width, $l_{U,\bot}$ is the perpendicular UNM width, $n_{peak}$ is the peak number density for CNM which is restricted by the HI-H2 conversion, and $n_C$ is the mean density of the CNM. Notice that Eq.\ref{eq:aspect_UNM_compression} is useful in quantifying whether a particular feature is a dynamically stable CNM.

VDA can also determine whether the structures in channel maps are part of the same turbulent system. In \cite{spectrum}, it is shown that: (1) features that are magnetized and parallel to the magnetic fields tend to have a density spectral power slope of $-2$; (2) if the feature is not magnetized, the density power spectral slope is steeper and can be as most as $-11/3 \approx -3.7$ (See e.g. \cite{LP00,LP04}). By examining the power spectra of each individual regions in the {\it Cattail}, we can determine whether the structures are likely in the same turbulence system. VDA also demonstrates that when features are disjoint in the caustic map they are unlikely to be a part of the same turbulent system (See also \citealt({population}).

\section{Results}
\label{sec:results}

To show that the {\it Cattail} (1.1 kpc region) is likely not in the CNM (and therefore likely to be UNM), we establish that: (1) There is a significant turbulent velocity fluctuation in the {\it Cattail} region (2) the region is thermally dominant (3) It is not dynamically stable if it is CNM \cite{HO21}.

\subsection{VDA reveals the existence of Kolmogorov turbulence cascade in {\it Cattail}}
\begin{figure}
\includegraphics[width=0.49\textwidth]{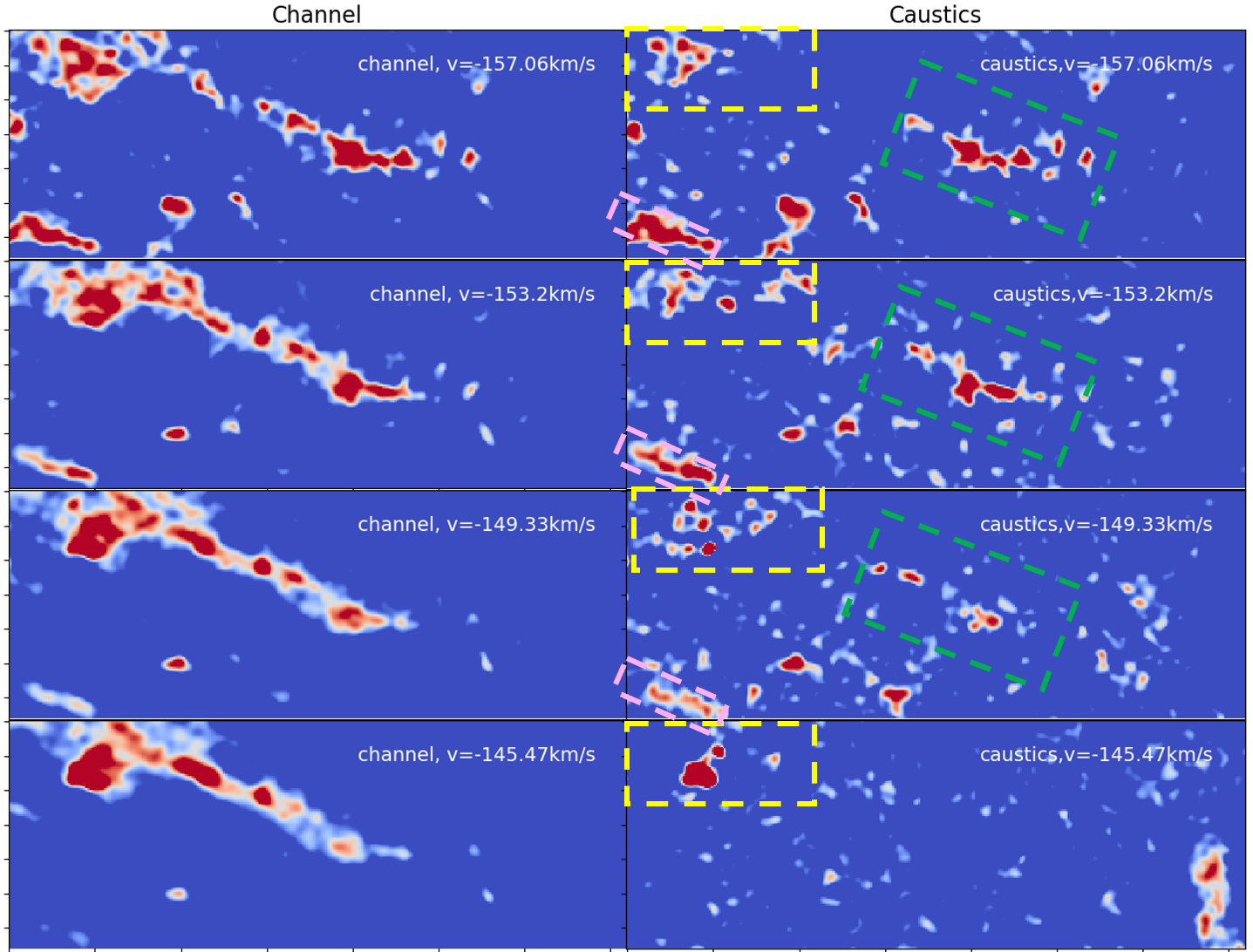}
\caption{A set of figures showing the fluctuations of velocity channels (left) and caustics (right) at least $0.8\sigma-3\sigma$ above the mean value for different selected velocities. Here we marked three spatially separate regions for our discussions in \S \ref{sec:spat}. The green region correspond to the tip of the {\it Cattail} (the 1.1 kpc one) that is discussed in the first half of \citep{Li21}. }
\label{fig:channeldisp}
\end{figure}

Since {\it Cattail} is at the edge of the spectral line, using the $\sigma-v$ diagram in analyzing the relative fluctuations across $v$ requires additional consideration (Yuen et. al in prep). Hence we can only use the $P_d/P_v$ diagram. The left of Fig.\ref{fig:PdPv} shows the spectrum ratio of the density and velocity fluctuations $P_d/P_v$ in the "grand" (i.e. 5kpc) {\it Cattail} region for the 4 selected thin\footnote{\cite{LP00} defines "thin" velocity channel when the channel width $\Delta v$ is smaller than the spatial velocity dispersion $\delta v(R)$.} velocity slices as shown in Fig.\ref{fig:channeldisp}. We can see from this figure that the intermediate and small scale fluctuations in channel maps do indeed correspond to the velocity fluctuations since $P_d/P_v\ll 1$ roughly at $0.3-2.0$ kpc. This indicates a strong turbulent imprint in the {\it Cattail} region over a range of velocity slices, especially for the smaller {\it Cattail} region (1.1 kpc). To show that this region is also velocity dominant, on the right of the Fig. \ref{fig:PdPv} we perform the same $P_d/P_v$ analysis by focusing only on the green box. Notice that the velocity fluctuation at $v=-145$ is almost zero visually (See the lower right panel of Fig.\ref{fig:channeldisp}), therefore this line is excluded. We can see from the right of Fig. \ref{fig:PdPv} that the entire "smaller" {\it Cattail} region is velocity dominant, meaning turbulence contributes significantly to the linewidth. This contribution is roughly $3-5 km/s$ \citep{kk16}\footnote{The $3-5 km/s$ contribution of turbulence to the linewidth is just a rough estimate. It is a well-known that turbulence linewidth does not significantly vary in multiphase media (\cite{KH17}. The lower estimate of $3km/s$ comes from \cite{kk16} by the full width half maximum of the decomposed CNM. While the $5km/s$ estimate came from \cite{kh18} where the authors are fitting the CNM multi-Gaussian spectral line. This estimate has also been verified numerically. See Appendix}.

In theory one can determine the strength of the thermal broadening by inspecting the channel map, however showing that the thermal broadening exist does not mean that turbulence fluctuations exist in channel maps (See, counterexamples from \citealt{Clark19} and also \citealt{reply19}). Our result in this subsection shows that the turbulence fluctuations indeed exist and follows the expected Kolmogorov cascade (See also \cite{spectrum}).

\begin{figure}
\includegraphics[width=0.49\textwidth]{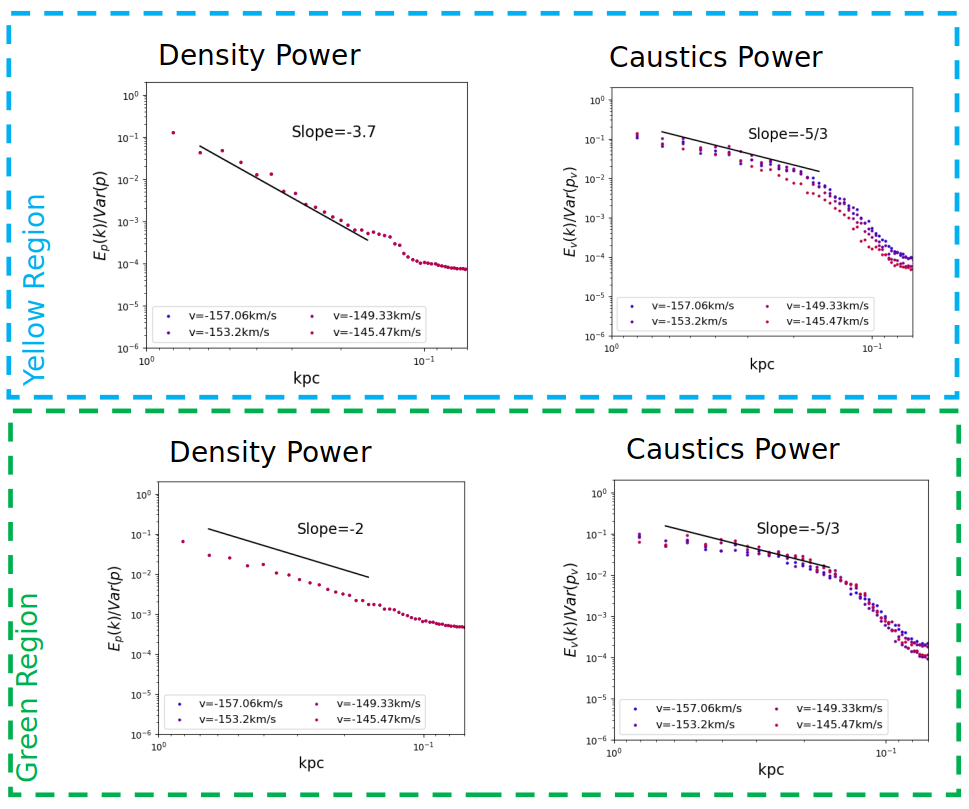}
\caption{Four figures showing the variation of the normalized power spectra for the 4 selected velocity slices as a function of length scales (in kpc). From top left: $p_d$ for the yellow region; top right: $p_v$ for yellow region; lower left: $p_d$ for the green region; lower right: $p_v$ for the green region.}
\label{fig:powerspec}
\end{figure}

\begin{figure*}
\includegraphics[width=0.49\textwidth]{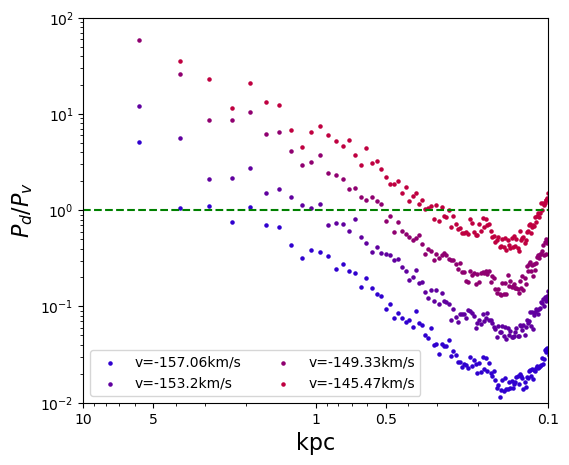}
\includegraphics[width=0.49\textwidth]{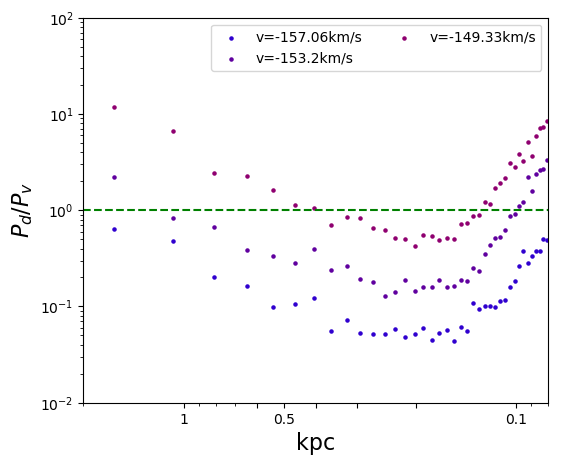}
\caption{(Left) A figure showing the spectrum ratio of the density and velocity fluctuations $P_d/P_v$ in the grand {\it Cattail} (5kpc) region for the 4 selected velocity slices as a function of length scales (in kpc) in Fig.\ref{fig:channeldisp}. (Right) The $P_d/P_v$ analysis for the tip of the {\it Cattail} (1.1kpc) region. Notice that we exclude $v=-145km/s$ for this panel because there are no noticeable caustics contribution as we see in Fig .\ref{fig:channeldisp}.}
\label{fig:PdPv}
\end{figure*}

\subsection{VDA confirms the strong thermal broadening signal}

We apply Eq.\ref{eq:VDA_ld2} and visually inspect the fluctuations of thin channels near the {\it Cattail} region (in this instance the channels are thin because $\Delta v \sim 1.288 km/s$ (the slices in PPV) < $v_{turb}$ ($3-5km/s$) (see \cite{LP00,LP04,LY18a} for a precise definition of thin channels) near the {\it Cattail} region. If the feature is not subject to strong thermal broadening we expect to see a change of channel map features over neighboring channels due to caustics fluctuations. Fig.\ref{fig:channeldisp} shows the fluctuations of HI channel map and caustics map side by side via VDA with fluctuations above mean for at least $0.8 \sigma$. This guarantees the noise contribution will be eliminated. According to VDA, only when the system is thermally dominant will the channel maps look more or less the same across different velocities. In this scenario, VDA will remove the contributions due to thermal broadening and reveal the real turbulent fluctuations of the system in the caustic map. We can easily recognize that the {\it Cattail} features in channel maps stay pretty much the same over 10 km/s, thus indicating the presence of strong thermal broadening. 

\subsection{Force Balancing Model shows that {\it Cattail} is dynamically unstable if CNM}

To proceed with applying the Force Balancing Model to {\it Cattail} we have to estimate the parameters in Eq.\ref{eq:aspect_UNM_compression} using :\\
\linebreak
{\noindent\bf $M_s$}: Using our estimation on the thermal linewidth from \S \ref{sec:spat}, we see that $M_s\approx \sqrt{3} v_{turb}/c_s \approx 1$.\\
\linebreak
{\noindent\bf $M_A$}: We can utilize a simple scaling relation to approximate the $M_A$ at $22kpc$ using the $M_A$ estimates of $\sim 0.5$ from nearby CNM \citep{curvature}: $M_A(R=22kpc) \sim \sqrt{\frac{n_H(R=22kpc)}{n_H(R=0kpc)}} M_A(R=0kpc)$ (using ${n_H}$ from \cite{Li21}). Given that $n_H$ in R=0kpc is $\sim 10cm^{-3}$, $M_A$ at $22kpc$ is roughly $\sqrt{0.03/10}\times 0.5 \sim 0.02$, which agrees with the visual absence of the structure curvature in the {\it Cattail} region \citep{curvature}.\\
\linebreak
{\noindent\bf $l_\perp/l_{U,\perp}$}:  The factor in the bracket depends on two parameters. The $l_\perp/l_{U,\perp}$ parameter can be approximated by the {\it volume} fraction of CNM and UNM similar to \cite{HO21} as 1 (which corresponds to $CNM:UNM \sim 1:8$ in volume fractions). \\\linebreak
{\noindent\bf $n_{peak}/n_C$} The other parameter $n_{peak}/n_C$ can be estimated by the peak intensity value near the {\it Cattail} region ($1.7\times 10^{19} cm^{-2}$) over their "CNM" mean volume density estimates ($0.03 cm^{-3})$ times their estimated LOS distance ($200pc$) which is roughly 1. 

Summarizing these numbers, we have an estimate for the aspect ratio that is not much deviated from unity. According to \cite{HO21} this indicates that if {\it Cattail} is CNM, it is unlikely to be dynamically stable, i.e. it will be fragmented in the foreseeable future. In addition, in the whole {\it Cattail} (5 kpc) we observe a number of clumpy, circular features from visual inspections in velocity channels. These clumpy, spherical features are inconsistent with the usual expectation that the CNM is highly anisotropic and elongated along B-field.

\subsection{VDA also shows that {\it Cattail} consists of multiple systems and likely does not extend for 5 kpc as a single turbulent system} \label{sec:spat}
From the visual inspection of Fig.\ref{fig:channeldisp}, we can see that near the mean velocity of {\it Cattail} ($v\sim 151km/s$), the fluctuations of velocity caustics are actually spatially separated into a few groups. We draw their boundaries with rectangular boxes. Notice that the green region is the "smaller" {\it Cattail} that is defined in the first half of \cite{Li21} paper. These features seem to be connected in channel maps but are disjoint in all caustics maps. The disconnection of features are also seen if one inspects the velocity channels $v<-158 km/s$. We also analyze the velocity caustics power spectrum for the yellow and green regions labelled in Fig.\ref{fig:channeldisp}. The left of Fig.\ref{fig:powerspec} shows the power spectra for both $p_d$ and $p_v$ in the green and yellow regions. We can see that the slope of their density power spectra are different: The green region has a density spectral slope of $-2$, which indicates the feature is magnetized and parallel to the magnetic field \citep{spectrum,population}, while in the yellow region we observe a density power spectral slope of $\sim -3.7$, which is a signature of weakly magnetized turbulence. This difference suggests that the yellow and the green region are not in the same turbulent system. As a result it is very unlikely that {\it Cattail} can be considered as a coherent region 5 kpc long.

\section{Quantitative argument why the tip of {\it Cattail} is unlikely cold}
\label{sec:qual}

By combining the facts in \S \ref{sec:results}, we see that the {\it Cattail} is unlikely to be CNM. In this section we shall give quantitative arguments that is consistent with our results. \bigskip

{\noindent \bf {Discrepancy of linewidth contribution even with the presence of galactic rotation curve}} We discussed in the introduction that even considering the galactic rotation contributions, it is very unlikely that the {\it Cattail} at $T=200K$ will have a linewidth of 12 km/s, unless there are other physical processes occurring in {\it Cattail}, such as outflow velocities along the line of sight, supernova/shock driven bubbles etc. However, it is hard to imagine that these physical processes are happening at length scales of 1 kpc (or 5 kpc, depending on the definition of {\it Cattail}) at 22 kpc away from the galactic center given that the region seems to be self-isolated. It is therefore more natural to assume that the contributions of linewidth are only from turbulence, galactic shear and thermal broadening. 

Shear galactic rotation at 22 kpc contributes roughly 3 km/s to {\it Cattail's} linewidth.\footnote{In \cite{Li21} the line of sight depth of {\it Cattail} is estimated to be 200 pc. We choose the very upper estimate of a flat galactic rotation curve model of the Milky Way $v_{rotation}< 300km/s$. For a 200 pc object, this will naturally introduce a total amount of {\bf internal} shear of $300 km/s \times 200pc/22kpc < 3km/s$. We also assume that the shear contribution is all projected along the line of sight. This would constitute the upper estimate of shear contribution. The choice of upper estimate is to minimize the thermal width estimated in this section to make it colder.} Turbulence also contributes roughly 3 - 5 km/s to the linewidth (See \S 3.1). If there are no contributions from other physical processes, then the thermal width is estimated to be $4-6 km/s$. This will corresponds to a temperature of 940K ($\bar{c}_s=4km/s$) to 2130 K ($\bar{c}_s=6km/s$)\footnote{ $T \sim (v/13km/s)^2 \times 10,000 K$}, consistent with the temperature range of the UNM\footnote{Noticing that the definition of CNM in terms of temperature changes as a function of the cooling and heating function. However CNM rarely will be at T=1000K, see \citealt{Wolfire03}.}.However in \S \ref{sec:discussions} we emphasize that we cannot solely use the temperature to determine the phases. More importantly, we also do not exclude the possibility that the region might be a mixture of warm and cold gases as proposed by \cite{K&D08}. \linebreak

{\noindent \bf {Density too small to be CNM}} At the outer parts of the galaxy it is hard to achieve the tight pressure balances suggested by \cite{F69}. Yet one can still assume that such a balance occurs and consider the properties of the unstable and cold phase with the density and the temperature quoted from \cite{Li21}. We find that these numbers can be roughly compared {\it order of magnitude wise} with the 18 kpc result form \cite{Wolfire03}. Although from \cite{KK09} we see that the mean density of the HI clouds at 22 kpc is of the order $0.01 cm^{-3}$, this density corresponds to the warm gases rather than the cold gases. The density of the CNM is at least 30-70 times that of WNM (e.g. compare Table 3 of \cite{Wolfire03} to that of Fig 4 of \cite{KK09}). Therefore a rough estimation of the CNM density in {\it Cattail} should be $n_{CNM}\sim 50\times n_{{\it Cattail}} \sim 1.5 cm^{-3}$. \linebreak

{\noindent \bf {Pressure too small to be CNM}} It is also apparent that the pressure estimate from the numbers given by \cite{Li21} is too low compared to \cite{Wolfire03}. By a simple computation, $P/k_B \sim 0.03 \times 200 \sim 6 K cm^{-3}$. Yet in \cite{Wolfire03}, the pressure at 18kpc is roughly $0.025 \times 8000 \sim 200 K cm^{-3}$. Assuming proper interpolations, the estimation of the pressure of {\it Cattail}, if it is CNM, is still two orders of magnitude lower than that of Fig 9 of \cite{Wolfire03} and Fig 12 of \cite{KK09}.

\section{Discussions}
\label{sec:discussions}

\subsection{The Importance of understanding the UNM}

The definition of CNM is not well defined despite being used for many years in the literature. Employing varying definitions (200-700K), the UNM regime (ranging from 200K to 600K, see also \cite{Murray18}) is often miss-categorized as the "cold phase". This temperature-based classification of CNM often lead to erroneous conclusions in terms of its physical properties, e.g. an inadmissible aspect ratio for CNM \cite{HO21}. One should therefore take caution that any definition of CNM should be given by the respective phase diagram defined by the balances of heating and cooling rates as opposed to a specified temperature threshold.


The UNM plays a significant role in many interstellar processes. Having high mass and volume fraction (~40\% for both, \cite{kk16}), UNM controls the formation and shaping of CNM features (See \cite{HO21}) by providing additional stability to the CNM at the cost of smaller aspect ratio. Due to strong self-absorption (when cold HI in the foreground absorbs the emission of warmer HI in the background) and quick H2 conversion lifetime \cite{BS16}, it was recently suggested \cite{SF20} that the HI emission map might actually not detecting all of the CNM. Moreover, UNM has generally longer aspect ratio than the embedded CNM \cite{HO21}, which is believed to be the result of the grand cascade despite \citeauthor{GS95} scaling not being relevant during the phase transition. Turbulence also plays an important role in maintaining the existence and stability of the UNM \citep{lifetime}. Indeed, if magnetized turbulence is not involved in the formation of UNM, the observed fraction of UNM will be significantly lower and UNM will quickly dissipate. Simulations (\citealt{AH05,KH17,HO21,Murray18,lifetime})) showed that the UNM with adequate turbulent injection will display a mass/volume fraction of 0.4, and is consistent with observational findings (\cite{kh18} mass fraction $\sim$ volume fraction $\sim$ 0.4). Observing a huge, linear UNM on the sky (e.g. 1.1 kpc {\it Cattail}) would be a huge step in understanding the actual dynamics of UNM and how it impacts the formation of CNM especially with better inferometric instruments like FAST.

\subsection{Kolmogorov cascade observationally verified in {\it Cattail}}

In this paper, We showed that turbulence is an essential component in {\it Cattail}. In fact, from our analysis, we see that a grand Kolmogorov $-5/3$ cascade can be recovered via caustic maps, despite the channel maps not having such power scaling. It has been strongly postulated that turbulence cascades from very large scale down to cores and filaments, but the only observational proof aside from the Larson's classic paper (1981) is the electron density power spectra provided by \cite{Chepurnov2010ExtendingData} and \cite{Armstrong1995ElectronMedium}). However, the most important and dynamical length scales, 0.01 pc to 10 kpc, where the entire dynamics of HI and some of H2, are not covered, and the evidence of Kolmogorov cascade is subject to outlier effect.  \cite{spectrum} recently showed a grand Kolmogorov $-5/3$ caustics cascade from >100 pc down to 0.01 pc using data from multiple tracers. Our study complements the study \citep{spectrum} that even at the length scale of {\it Cattail} ($\sim$ kpc), we can still recover the Kolmogorov $-5/3$ slope from caustics power spectrum, indicating the relevance of Kolmogorov-type cascade (including \cite{GS95}) in multiphase media.

\subsection{A cold neutral media candidate discovered in VDA}

From our analysis we observe a much smaller and compact region, which we labelled in Fig.\ref{fig:channeldisp} in a pink rectangle, that is not removed by VDA. We recognize that this smaller region has a much smaller linewidth ($\langle v\rangle \sim 160 km/s, \Delta v \sim 5 km/s$), consistent with the linewidth estimation of the CNM in the nearby galactic arm. This smaller filament is anisotropic, and has a rough aspect ratio of 3-5 (which is plausible since the permitted aspect ratio is larger for stable CNM as aspect ratio $\propto M_s^2$ ), consistent with the \cite{HO21} restriction on stable CNM aspect ratio.

\section{Conclusion}
\label{sec:conclusion}
In this paper, we analyze the {\it Cattail} with the VDA and Force Balancing Model. We provide a simple analysis to determine the turbulent,thermal and dynamical properties of the {\it Cattail} region.. We find that:
\begin{enumerate}
    \item The {\it Cattail} is likely to be in UNM (\S 3, \S 4).
    \item The {\it Cattail} region is turbulent (\S 3.1, Fig \ref{fig:PdPv})
    \item The clumpy feature in {\it Cattail} is explained by the instability happening due to the existence of UNM (\S 3.3)
    \item The {\it Cattail} consists of several turbulent system and likely does not extend 5 kpc as a single turbulent system (\S 3.4 Fig.\ref{fig:channeldisp}) 
    \item The caustics power spectra for different regions of{\it Cattail} exhibit the Kolmogorov $-5/3$ law, which is consistent with \cite{spectrum} (\S 3.2,5.2 Fig.\ref{fig:powerspec})

\end{enumerate}

The first finding of a massive UNM filament will likely help us to understand the properties of multiphase HI more clearly in the near future.

{\noindent \bf Acknowledgment} K.H.Y., K.W.H. \& A.L. acknowledge the support the NSF AST 1816234, NASA TCAN 144AAG1967 and NASA ATP AAH7546. The numerical part of the research used resources of both Center for High Throughput Computing (CHTC) at the University of Wisconsin and National Energy Research Scientific Computing Center (NERSC), a U.S. Department of Energy Office of Science User Facility operated under Contract No. DE-AC02-05CH11231, as allocated by TCAN 144AAG1967. K.H.Y, K.W.H acknowledges Chi Yan Law (Chalmers) for discussions.


\bsp	

\appendix

\section{turbulent line width for three phase medium in numerical simulation}

\begin{table}[]
\label{table: simulation} 
\begin{tabular}{@{}lllll@{}}
\toprule
                     & \multicolumn{3}{c}{\begin{tabular}[c]{@{}c@{}}Weak field case\\ n\_0 = 3\\ B\_0 = 1\end{tabular}}  \\ \midrule
                     & WNM                            & UNM                            & CNM                                                  \\
$M_s$                   & 0.7                            & 1.6                            & 6.0                                \\
$M_A$                   & 0.8                            & 1.6                            & 2.2                               \\
Mass filling factor $M_{C/U/W}$ & 0.08                           & 0.28                           & 0.63                              \\
Volume filling factor $V_{C/U/W}$ & 0.61                           & 0.31                           & 0.070                           \\ \bottomrule
\end{tabular}
\caption{Simulation parameters and the basic statistics for each phase in our multiphase simulation, from \citep{VDA,HO21}.}
\end{table}

\begin{figure}
\label{fig:phase_velocity}
\includegraphics[width=0.49\textwidth]{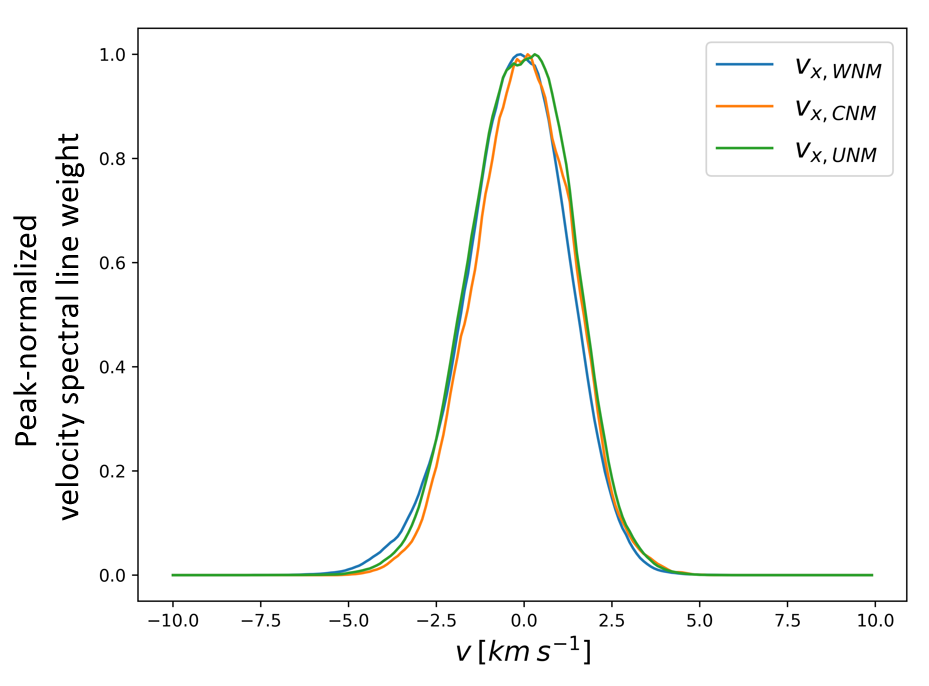}
\caption{The velocity spectral lines for three phases in multi-phase simulation we employed in \citep{VDA,HO21}, where the x-axis shows the velocity range in the unit of $\protect km s^{-1}$ and the y-axis shows the peak-normalized velocity spectral line weight ranging from 0 to 1.}
\end{figure}

At \S 3, we discussed the turbulent contribution to the line-width of {\it Cattail}. While the temperature of HI gas varies for two orders of magnitude in multi-phase medium, its velocity distribution keeps constant throughout the phase transition. This can be shown using the multi-phase MHD simulation that we used in  \cite{VDA} and \cite{HO21}. Fig. \ref{fig:phase_velocity} shows the velocity distribution of three phases using the "weak field" simulation in \cite{VDA} and \cite{HO21}. One can see that the three phases share a highly similar distribution with approximately the same FWHM, suggesting that we can use the same turbulent line width for all phases in our analysis (\S \ref{sec:qual}).

\label{lastpage}

@ARTICLE{2013A&A...558A..33A,
       author = {{Astropy Collaboration} and {Robitaille}, Thomas P. and
         {Tollerud}, Erik J. and {Greenfield}, Perry and {Droettboom}, Michael and
         {Bray}, Erik and {Aldcroft}, Tom and {Davis}, Matt and
         {Ginsburg}, Adam and {Price-Whelan}, Adrian M. and
         {Kerzendorf}, Wolfgang E. and {Conley}, Alexander and {Crighton}, Neil and
         {Barbary}, Kyle and {Muna}, Demitri and {Ferguson}, Henry and
         {Grollier}, Fr{\'e}d{\'e}ric and {Parikh}, Madhura M. and
         {Nair}, Prasanth H. and {Unther}, Hans M. and {Deil}, Christoph and
         {Woillez}, Julien and {Conseil}, Simon and {Kramer}, Roban and
         {Turner}, James E.~H. and {Singer}, Leo and {Fox}, Ryan and
         {Weaver}, Benjamin A. and {Zabalza}, Victor and {Edwards}, Zachary I. and
         {Azalee Bostroem}, K. and {Burke}, D.~J. and {Casey}, Andrew R. and
         {Crawford}, Steven M. and {Dencheva}, Nadia and {Ely}, Justin and
         {Jenness}, Tim and {Labrie}, Kathleen and {Lim}, Pey Lian and
         {Pierfederici}, Francesco and {Pontzen}, Andrew and {Ptak}, Andy and
         {Refsdal}, Brian and {Servillat}, Mathieu and {Streicher}, Ole},
        title = "{Astropy: A community Python package for astronomy}",
      journal = {\aap},
     keywords = {methods: data analysis, methods: miscellaneous, virtual observatory tools, Astrophysics - Instrumentation and Methods for Astrophysics},
         year = "2013",
        month = "Oct",
       volume = {558},
          eid = {A33},
        pages = {A33},
          doi = {10.1051/0004-6361/201322068},
archivePrefix = {arXiv},
       eprint = {1307.6212},
 primaryClass = {astro-ph.IM},
       adsurl = {https://ui.adsabs.harvard.edu/abs/2013A&A...558A..33A},
      adsnote = {Provided by the SAO/NASA Astrophysics Data System}
}

@ARTICLE{1996A&AS..117..393B,
       author = {{Bertin}, E. and {Arnouts}, S.},
        title = "{SExtractor: Software for source extraction.}",
      journal = {\aaps},
     keywords = {METHODS: DATA ANALYSIS, TECHNIQUES: IMAGE PROCESSING, GALAXIES: PHOTOMETRY},
         year = "1996",
        month = "Jun",
       volume = {117},
        pages = {393-404},
          doi = {10.1051/aas:1996164},
       adsurl = {https://ui.adsabs.harvard.edu/abs/1996A&AS..117..393B},
      adsnote = {Provided by the SAO/NASA Astrophysics Data System}
}

@ARTICLE{2018AJ....156...82C,
       author = {{Cloutier}, Ryan and {Doyon}, Ren{\'e} and {Bouchy}, Francois and
         {H{\'e}brard}, Guillaume},
        title = "{Quantifying the Observational Effort Required for the Radial Velocity Characterization of TESS Planets}",
      journal = {\aj},
     keywords = {methods: analytical, planets and satellites: detection, planets and satellites: fundamental parameters, techniques: radial velocities, Astrophysics - Earth and Planetary Astrophysics},
         year = "2018",
        month = "Aug",
       volume = {156},
       number = {2},
          eid = {82},
        pages = {82},
          doi = {10.3847/1538-3881/aacea9},
archivePrefix = {arXiv},
       eprint = {1807.01263},
 primaryClass = {astro-ph.EP},
       adsurl = {https://ui.adsabs.harvard.edu/abs/2018AJ....156...82C},
      adsnote = {Provided by the SAO/NASA Astrophysics Data System}
}

@ARTICLE{2015ApJ...805...23C,
       author = {{Corrales}, Lia},
        title = "{X-Ray Scattering Echoes and Ghost Halos from the Intergalactic Medium: Relation to the Nature of AGN Variability}",
      journal = {\apj},
     keywords = {accretion, accretion disks, dust, extinction, quasars: general, intergalactic medium, X-rays: ISM, Astrophysics - High Energy Astrophysical Phenomena, Astrophysics - Astrophysics of Galaxies},
         year = "2015",
        month = "May",
       volume = {805},
       number = {1},
          eid = {23},
        pages = {23},
          doi = {10.1088/0004-637X/805/1/23},
archivePrefix = {arXiv},
       eprint = {1503.01475},
 primaryClass = {astro-ph.HE},
       adsurl = {https://ui.adsabs.harvard.edu/abs/2015ApJ...805...23C},
      adsnote = {Provided by the SAO/NASA Astrophysics Data System}
}

@ARTICLE{2013RMxAA..49..137F,
       author = {{Ferland}, G.~J. and {Porter}, R.~L. and {van Hoof}, P.~A.~M. and
         {Williams}, R.~J.~R. and {Abel}, N.~P. and {Lykins}, M.~L. and
         {Shaw}, G. and {Henney}, W.~J. and {Stancil}, P.~C.},
        title = "{The 2013 Release of Cloudy}",
      journal = {\rmxaa},
     keywords = {atomic processes, galaxies: active, methods: numerical, molecular processes, radiation mechanisms: general, Astrophysics - Galaxy Astrophysics, Astrophysics - Cosmology and Extragalactic Astrophysics, Astrophysics - Instrumentation and Methods for Astrophysics},
         year = "2013",
        month = "Apr",
       volume = {49},
        pages = {137-163},
archivePrefix = {arXiv},
       eprint = {1302.4485},
 primaryClass = {astro-ph.GA},
       adsurl = {https://ui.adsabs.harvard.edu/abs/2013RMxAA..49..137F},
      adsnote = {Provided by the SAO/NASA Astrophysics Data System}
}

@INPROCEEDINGS{1989BAAS...21..780H,
       author = {{Hanisch}, R.~J. and {Biemesderfer}, C.~D.},
        title = "{T$_{E}$X and LAT$_{E}$X Macro Definition Files for Astronomical Publications}",
    booktitle = {\baas},
         year = "1989",
        month = "Mar",
        pages = {780},
       adsurl = {https://ui.adsabs.harvard.edu/abs/1989BAAS...21..780H},
      adsnote = {Provided by the SAO/NASA Astrophysics Data System}
}

@BOOK{lamport94,
       author = {{Lamport}, L.},
        title = "{LaTeX: A Document Preparation System}",
    publisher = {Addison-Wesley Professional},
	 year = "1994",
      edition = {2},
	 isbn = {0201529831}
}

@ARTICLE{2018ApJ...868L..33L,
       author = {{Li}, Leping and {Zhang}, Jun and {Peter}, Hardi and
         {Chitta}, Lakshmi Pradeep and {Su}, Jiangtao and {Song}, Hongqiang and
         {Xia}, Chun and {Hou}, Yijun},
        title = "{Quasi-periodic Fast Propagating Magnetoacoustic Waves during the Magnetic Reconnection Between Solar Coronal Loops}",
      journal = {\apj},
     keywords = {magnetic reconnection, plasmas, Sun: corona, Sun: UV radiation, waves, Astrophysics - Solar and Stellar Astrophysics},
         year = "2018",
        month = "Dec",
       volume = {868},
       number = {2},
          eid = {L33},
        pages = {L33},
          doi = {10.3847/2041-8213/aaf167},
archivePrefix = {arXiv},
       eprint = {1811.08553},
 primaryClass = {astro-ph.SR},
       adsurl = {https://ui.adsabs.harvard.edu/abs/2018ApJ...868L..33L},
      adsnote = {Provided by the SAO/NASA Astrophysics Data System}
}

@ARTICLE{2016AJ....152...41P,
       author = {{Pr{\v{s}}a}, Andrej and {Harmanec}, Petr and {Torres}, Guillermo and
         {Mamajek}, Eric and {Asplund}, Martin and {Capitaine}, Nicole and
         {Christensen-Dalsgaard}, J{\o}rgen and {Depagne}, {\'E}ric and
         {Haberreiter}, Margit and {Hekker}, Saskia},
        title = "{Nominal Values for Selected Solar and Planetary Quantities: IAU 2015 Resolution B3}",
      journal = {\aj},
     keywords = {planets and satellites: fundamental parameters, standards, stars: fundamental parameters, stars: general, Sun: fundamental parameters, Astrophysics - Solar and Stellar Astrophysics, Astrophysics - Earth and Planetary Astrophysics, Astrophysics - Instrumentation and Methods for Astrophysics},
         year = "2016",
        month = "Aug",
       volume = {152},
       number = {2},
          eid = {41},
        pages = {41},
          doi = {10.3847/0004-6256/152/2/41},
archivePrefix = {arXiv},
       eprint = {1605.09788},
 primaryClass = {astro-ph.SR},
       adsurl = {https://ui.adsabs.harvard.edu/abs/2016AJ....152...41P},
      adsnote = {Provided by the SAO/NASA Astrophysics Data System}
}

@ARTICLE{2011ApJS..197...31S,
       author = {{Schwarz}, Greg J. and {Ness}, Jan-Uwe and {Osborne}, J.~P. and
         {Page}, K.~L. and {Evans}, P.~A. and {Beardmore}, A.~P. and
         {Walter}, Frederick M. and {Helton}, L. Andrew and
         {Woodward}, Charles E. and {Bode}, Mike and {Starrfield}, Sumner and
         {Drake}, Jeremy J.},
        title = "{Swift X-Ray Observations of Classical Novae. II. The Super Soft Source Sample}",
      journal = {\apjs},
     keywords = {novae, cataclysmic variables, ultraviolet: stars, X-rays: stars, Astrophysics - Solar and Stellar Astrophysics, Astrophysics - High Energy Astrophysical Phenomena},
         year = "2011",
        month = "Dec",
       volume = {197},
       number = {2},
          eid = {31},
        pages = {31},
          doi = {10.1088/0067-0049/197/2/31},
archivePrefix = {arXiv},
       eprint = {1110.6224},
 primaryClass = {astro-ph.SR},
       adsurl = {https://ui.adsabs.harvard.edu/abs/2011ApJS..197...31S},
      adsnote = {Provided by the SAO/NASA Astrophysics Data System}
}

@ARTICLE{2014ApJ...793..127V,
       author = {{Vogt}, Fr{\'e}d{\'e}ric P.~A. and {Dopita}, Michael A. and
         {Kewley}, Lisa J. and {Sutherland}, Ralph S. and
         {Scharw{\"a}chter}, Julia and {Basurah}, Hassan M. and {Ali}, Alaa and
         {Amer}, Morsi A.},
        title = "{Galaxy Emission Line Classification Using Three-dimensional Line Ratio Diagrams}",
      journal = {\apj},
     keywords = {galaxies: abundances, galaxies: general, galaxies: Seyfert, galaxies: starburst, H II regions, ISM: lines and bands, Astrophysics - Astrophysics of Galaxies},
         year = "2014",
        month = "Oct",
       volume = {793},
       number = {2},
          eid = {127},
        pages = {127},
          doi = {10.1088/0004-637X/793/2/127},
archivePrefix = {arXiv},
       eprint = {1406.5186},
 primaryClass = {astro-ph.GA},
       adsurl = {https://ui.adsabs.harvard.edu/abs/2014ApJ...793..127V},
      adsnote = {Provided by the SAO/NASA Astrophysics Data System}
}


\begin{thebibliography}{}
\providecommand\natexlab[1]{#1}
\providecommand\JournalTitle[1]{#1}
%
\footnotesize

\bibitem[Armstrong et al.(1995)]{Armstrong1995ElectronMedium}
Armstrong, J.~W., Rickett, B.~J., \& Spangler, S.~R. 1995,\href{http://dx.doi.org/10.1086/175515}{\JournalTitle{The Astrophysical Journal}, 443, 209}

\bibitem[Arnold et al. (1985)]{AR85}Arnold, V.I., Varchenko, A. and Gusein-Zade, S.M., Singularities of
Differentiable Maps: Volume I: The Classification of Critical
Points Caustics and Wave Fronts, 1985, Birkhauser, Bosto

\bibitem[Audit \& Hennebelle (2005)]{AH05} Audit, E., \& Hennebelle, P., 2005, A\&A, 433, 1

\bibitem[Andr$\tilde{e}$(2017)]{A17} Andr$\tilde{e}$, P., 2017, 	arXiv:1710.01030


\bibitem[Bialy \& Sternberg (2016)]{BS16} Bialy S., Sternberg A., 2016, \apj, 822, 83. doi:10.3847/0004-637X/822/2/83

\bibitem[Beuther et al. (2016)]{B16} Beuther H., Bihr S., Rugel M., Johnston K., Wang Y., Walter F., Brunthaler
A., et al., 2016, A\&A, 595, A32. doi:10.1051/0004-6361/201629143

\bibitem[{Chepurnov \& Lazarian (2010)}]{Chepurnov2010ExtendingData}
Chepurnov, A., \& Lazarian, A. 2010,  \href{http://dx.doi.org/10.1088/0004-637X/710/1/853}{\JournalTitle{The
 Astrophysical Journal, Volume 710, Issue 1, pp. 853-858 (2010).}, 710, 853}

\bibitem[Clark et al.(2014)]{Clark14} Clark, S.~E., Peek, J.~E.~G., \& Putman, M.~E.\ 2014, \apj, 789, 82 

\bibitem[Clark et al.(2015)]{Clark15} Clark, S.~E., Hill, J.~C., Peek, J.~E.~G., Putman, M.~E., \& Babler, B.~L.\ 2015, Physical Review Letters, 115, 241302 

\bibitem[Clark et al. (2019)]{Clark19} Clark, S. E., Peek, J. E. G., \& Miville-Deschenes, M.-A. 2019, \apj, 874, 171

\bibitem[Clark \& Hensley (2019)]{CnH19} Clark S. E., Hensley B. S., 2019, ApJ, 887, 136. doi:10.3847/1538-
4357/ab5803

\bibitem[Crutcher \& Heiles (2003)]{CH00} Crutcher R., Heiles C., Troland T. (2003) Observations of Interstellar Magnetic Fields. In: Falgarone E., Passot T. (eds) Turbulence and Magnetic Fields in Astrophysics. Lecture Notes in Physics, vol 614. Springer, Berlin, Heidelberg. \url{https://doi.org/10.1007/3-540-36238-X_6}

\bibitem[Field et al. (1969)]{F69} Field, G. B., Goldsmith, D. W., Habing, H. J. 1969, ApJ., 155, L149

\bibitem[{Goldreich \& Sridhar (1995)}]{GS95} Goldreich, P., Sridhar, S. 1995, \href{http://dx.doi.org/10.1086/174600}{\JournalTitle{The Astronomical Journal}, 438, 763} 

\bibitem[Ho et al. (2021a)]{HO21} K. W. Ho, K. H. Yuen, Lazarian, A., 2021a, submitted to \mnras

\bibitem[Ho et al. (2021b)]{lifetime} K. W. Ho, K. H. Yuen, Lazarian, A., 2021b, submitted to \mnras

\bibitem[Ho et al. (2022)]{population} K. W. Ho, K. H. Yuen, Lazarian, A., in preparation


\bibitem[Heiles, C (2001)]{H01} Heiles, C., 2001, \apj 551, 105

\bibitem[Kalberla \& Dedes (2008)]{K&D08} Kalberla, P.M.W., \& Dedes, L., 2008, A\&A 487, 951

\bibitem[Kalberla \& Kerp (2009)]{KK09} Kalberla, P. M. W., \& Kerp, J. 2009, ARA\&A, 47, 27

\bibitem[Kalberla et al.(2016)]{kk16} Kalberla, P.~M.~W., Kerp, J., Haud, U., et al.\ 2016, \apj, 821, 117 

\bibitem[Kalberla \& Haud(2015)]{kh15} Kalberla P. M. W., Haud U., 2015, A\&A, 578, A78. doi:10.1051/0004-
6361/201525859

\bibitem[Kalberla \& Haud(2018)]{kh18} Kalberla P. M. W., Haud U., 2018, A\&A, 619, A58. doi:10.1051/0004-6361/201833146

\bibitem[Kalberla \& Haud(2019)]{kh2019} Kalberla, P.~M.~W., \& Haud, U.\ 2019, \aap, 627, A112 


\bibitem[Kritsuk et al. (2017)]{KH17} Kritsuk, A. G., Ustyugov, S. D., Norman, M. L. 2017 New J. Phys. 19 065003

\bibitem[Kandel et al.(2016)]{KLP16} Kandel, D., Lazarian, A., \& Pogosyan, D.\ 2016, \mnras, 461, 1227 

\bibitem[Kandel et al.(2017a)]{KLP17a} Kandel, D., Lazarian, A., \& Pogosyan, D.\ 2017, \mnras, 464, 3617 

\bibitem[Kandel et al.(2017b)]{KLP17b} Kandel, D., Lazarian, A., \& Pogosyan, D.\ 2017, \mnras, 470, 3103 

\bibitem[Li et al.(2021)]{Li21} Li, C., Qiu, K., Hu, B., et al., 2021, \apjl, 918:L2

\bibitem[Lazarian \& Pogosyan(2000)]{LP00} Lazarian, A., \& Pogosyan, D.\ 2000, \apj, 537, 720 

\bibitem[Lazarian \& Pogosyan(2004)]{LP04} Lazarian, A., \& Pogosyan, D.\ 2004, \apj, 616, 943 

\bibitem[Lazarian et al.(2017)]{Letal17} Lazarian, A., Yuen, K.~H., Lee, H., \& Cho, J.\ 2017, \apj, 842, 30 

\bibitem[Lazarian \& Yuen(2018)]{LY18a} Lazarian, A., \& Yuen, K. H. 2018, \apj, 853, 96

\bibitem[Lazarian et al.(2018)]{LYH18} Lazarian, A., Yuen, K.~H., Ho, K.~W., et al.\ 2018, \apj, 865, 46.

\bibitem[Lazarian et al.(2020)]{ch5} Lazarian, A., Yuen, K.~H., \& Pogosyan, D.\ 2020, arXiv:2002.07996, submitted to ApJ

\bibitem[Lazarian et al.(2021)]{ch9} Lazarian, A., Yuen, K.~H., \& Pogosyan, D.\ 2021, submitted to ApJ

\bibitem[Murray et al. (2018)]{Murray18} Murray, C. E., Stanimirovi$\tilde{c}$, Goss, W. M. (2018) \apj 238,14

\bibitem[Mro$\tilde{z}$ (2019)]{MR19} Mro$\tilde{z}$, P., Udalski, A., Skowron, D. M., et al., (2019) \apjl 870:L10

\bibitem[N.M. McClure-Griffiths et al. (2006)]{MG16} McClure-Griffiths, N. M., Dickey, J. M., Gaensler, B. M., Green, A. J., \& Haverkorn, M. 2006, ApJ, 652, 1339

\bibitem[Peek et al.(2018)]{P18} Peek, J.~E.~G., Babler, B.~L., Zheng, Y., et al.\ 2018, \apjs, 234, 2

\bibitem[Seifried et al. (2021)]{SF20} Seifried, D., Beauther, H., Walch, S. (2021), 	arXiv:2109.10917 

\bibitem[Soler et al. (2020)]{Soler20} Soler, J. D., Beuther, H., Syed, J., et al. (2020), A\&A, 642, A163

\bibitem[Wang et al. (2020)]{W20} Wang Y., Beuther H., Rugel M. R., Soler J. D., Stil J. M., Ott J., Bihr S., et. al, 2020, A\&A, 634, A83. doi:10.1051/0004-6361/201937095

\bibitem[Wolfire et al. (2003)]{Wolfire03} Wolfire M. G., McKee C. F., Hollenbach D., Tielens A. G. G. M., 2003, \apj, 587, 278

\bibitem[Xu et al.(2019)]{2019ApJ...878..157X} Xu, S., Ji, S., \& Lazarian, A.\ 2019, \apj, 878, 157 

\bibitem[Yuen \& Lazarian(2017a)]{YL17a} 
Yuen, K.~H., \& Lazarian, A.\ 2017, \apjl, 837, L24 

\bibitem[Yuen \& Lazarian(2017b)]{YL17b} 
Yuen, K.~H., \& Lazarian, A.\ 2017, arXiv:1703.03026 

\bibitem[Lazarian \& Yuen (2018)]{LY18} Lazarian, A., \& Yuen, K. H. 2018, \apj, 865, 59

\bibitem[Yuen et al.(2019)]{reply19} Yuen, K.~H., Hu, Y., Lazarian, A., \& Pogosyan, D.\ 2019, arXiv:1904.03173 


\bibitem[Yuen \& Lazarian (2020a)]{GA} Yuen, K.~H. \&  Lazarian, A., 2020, \apj, 898, 65


\bibitem[Yuen \& Lazarian(2020b)]{curvature} Yuen, K.~H., \& Lazarian, A.\ 2020, arXiv e-prints, arXiv:2002.01926

\bibitem[Yuen et al. (2021a)]{VDA} 
Yuen, K.~H.,Ho, K.~W.\& Lazarian, A., 2021, \apj

\bibitem[Yuen et al. (2021b)]{spectrum} 
Yuen, K.~H.,Ho, K.~W., Law, C.~Y, Chen,A. \& Lazarian, A., 2021, submitted


\end{thebibliography}
\end{document}